\documentclass[showpacs,amsmath,nofootinbib,amssymb,twocolumn, epsfig]{revtex4}
\usepackage{graphicx} 
\usepackage{dcolumn}
\usepackage{bm}

\usepackage{color}

\begin{document}

\title{Evolution of spherical overdensities in new agegraphic dark energy model}
\author{M. R. Setare }\email{rezakord@ipm.ir}
\affiliation{Department of Science, Campus of Bijar, University of Kurdistan, Bijar , Iran}
\author{F. Felegary}\email{falegari@azaruniv.ac.ir}
\author{F. Darabi}\email{f.darabi@azaruniv.ac.ir}
\affiliation{Department of Physics, Azarbaijan Shahid Madani University, Tabriz, 53714-161 Iran}

\date{\today}

\begin{abstract}
We study the structure
formation by investigating  the spherical collapse model in the context of new agegraphic dark energy model in flat FRW cosmology.
We compute the perturbational  quantities $g(a)$, $\delta_{c}(z_{c})$, $\lambda(z_{c})$,
$\xi(z_{c})$, $\Delta_{vir}(z_{c})$, $\log[\nu f(\nu)]$ and $\log[n(k)]$
for the new agegraphic dark energy model and compare the results with those
of EdS and $\Lambda CDM$ models.
We find that there is a dark energy dominated universe at low redshifts and a matter dominated universe at high redshifts in agreement with observations. Also, the size of structures, the overdense spherical region, and
the halo size in the new agegraphic  dark energy model are found smaller,
denser, and larger than those of EdS and $\Lambda CDM$ models.
We compare  our results with the results of tachyon scalar field
and holographic dark energy models.

\end{abstract}
\vspace{1cm}
\pacs{98.80.-k; 95.36.+x; 04.50.Kd.}
\maketitle

\section{Introduction}
The recent accelerated expansion of universe is one of the most significant cosmological discoveries over the last decade \cite{riess,perlmutter,De,sper}. In order to explain this unexpected behavior, many cosmological models have been proposed, some with the basis of modified gravity theories and
some with the basis of dark energy model. The cosmological constant is the
simplest candidate for dark energy but it suffers from cosmic coincidence and fine-tuning problems \cite{W,E}. The origin and nature of dark energy is still unknown. Thus, many different dark energy models
such as  holographic \cite{Li}, new agegraphic \cite{wei},
 phantom, quintom \cite{setare} and tachyon \cite{setare1,MR} have been proposed.
We know that the problem of  structure formation in the universe is a significant issue in theoretical cosmology. The spherical collapse model presented by Gott and Gunn \cite{Gott} is
the simplest structure formation  model.
In this model, a small spherical region is supposed subject to a homogeneous perturbation which is set in a homogeneous background universe. Also,
in the spherical collapse model, we confront with the important concepts such as virialization and turn-around.
The perturbation grows and quits the linear regime as time passes.
When  the radius $R$ becomes maximum, the perturbation stops expanding
and the Hubble flow decouples from the homogenous background, this is called turn-around. After this epoch, the perturbation starts
contracting. For a perfect pressureless matter and perfect spherical
symmetry, the perturbation collapses to a single point. However, since there
is hardly any perfect spherical symmetric overdensity in the universe, the
corresponding perturbation does not collapse to a single point. Finally, a virialized object
of a finite size is formed that is called Halo.
In addition, the evolution of structure growth have been investigated in different dark energy models such as: ghost \cite{malekjani}, tachyon \cite{tachyon}, chaplygin gas \cite{pace}, holographic \cite{holog}
and etc.\\
In this paper, we study the evolution of the growth
of overdense structures with respect to the dynamics of cosmic redshift or
scale factor. The dynamics of overdense structure depends on the expansion of universe and the dynamics of the background Hubble flow.
The spherical collapse model has been discussed thoroughly in Refs.\cite{fillmore,hoffman,ryden}.
In this work, we study the evolution of spherical overdensities in the new agegraphic dark energy model (NADE) and  compare our results with the results of EdS and $\Lambda CDM$ models. Also, we compare   our results with the results of tachyon scalar field
 model \cite{tachyon} and holographic dark energy model \cite{holog}.
\section{cosmology with new agegraphic dark energy model}
W know that the cosmological constant suffers from cosmic coincidence and fine-tuning problems known altogether as the cosmological constant problem. In general relativity, the space-time can be measured without any limit of accuracy. However, in quantum mechanics, the Heisenberg uncertainty relation imposes a limit of accuracy in these measurements \cite{wei}. Károlyházy and his collaborators
\cite{karo} constructed an interesting observation about the distance measurement
$t$ for Minkowski space-time  given by
\begin{equation}
\delta t=\lambda t_{p}^{\frac{2}{3}}t^{\frac{1}{3}}.\label{deltat}
\end{equation}
Here $\lambda$ is a dimensionless constant of order unity \cite{mazi,mazii}.
 In this work, we consider $\hbar=c=k_{B}=1$,  $l_{p}=t_{p}={m_{p}^{-1}}$ where
$l_{p}$, $t_{p}$ and $m_{p}$ are reduced Planck length, time and  mass, respectively.
Eq. (\ref{deltat}) together with the time-energy uncertainly relation
provides the possibility to estimate an energy density of the metric quantum fluctuations
of Minkowski space-time \cite{mazi,mazii}.
According to \cite{mazi,mazii}, with respect to Eq. (\ref{deltat}) a length
scale $t$ can be known with a maximum accuracy $\delta t$ determining thereby a minimal detectable cell
$\delta t^{3}\sim t_{p}^{2}t$ over a spatial region $t^{3}$. Such a cell expresses a minimal
detectable unit of space time over a given length scale $t$. If the age of Minkowski space time is $t$, then over a
spatial region with linear size $t$ there exists a minimal cell $\delta t^{3}$, whose energy cannot be smaller than
\cite{mazi,mazii}
\begin{equation}
E_{\delta t^{3}}\sim t^{-1},
\end{equation}
due to time-energy uncertainly relation.
Thus, the energy density of metric
quantum fluctuations of Minkowski space-time is given by \cite{mazi,mazii}
\begin{equation}
\rho_{q}\sim\frac{E_{\delta t^{3}}}{\delta t^{3}}\sim\frac{1}{t_{p}^{2}t^2}\sim\frac{m_{p}^{2}}{t^{2}}.\label{rho}
\end{equation}
With the  choice of age of the universe $T$, as the length scale in Eq. (\ref{rho}), one can obtain the agegraphic dark energy model as follows \cite{wei}

\begin{equation}
\rho_{q}=\frac{3n^{2}m_{p}^{2}}{T^{2}},\label{rho5}
\end{equation}
where $3n^{2}$ is of the order of unity and it
is introduced to parameterize some uncertainties such as the
effect of curved space-time and the species of quantum fields in the universe.
Since this model can not explain the matter dominated era, hence Wei and Cai 
proposed the new model that is called   new agegraphic dark energy model \cite{wei}.
In Eq. (\ref{rho}), one can choose the time scale to be the conformal time $\eta$  which is defined by $dt=a d\eta$. Therefore, the energy density
of new agegraphic dark energy is given by \cite{wei}
\begin{equation}
\rho_{q}=\frac{3n^{2}m_{p}^{2}}{\eta^{2}},\label{rhooo}
\end{equation}
where $3n^{2}$ is of order unity. 
The conformal time $\eta$ is given by
\begin{equation}
\eta\equiv\int\frac{dt}{da}=\int\frac{da}{a^{2}H}.\label{eta1}
\end{equation}
We consider a flat Friedmann-Robertson-Walker (FRW) universe containing new
agegraphic dark energy and pressureless matter. In a flat
FRW universe, the Friedmann equation is given by
\begin{equation}
H^{2}=\frac{1}{3m_{p}^{2}}(\rho_{q}+\rho_{m}),\label{friedman}
\end{equation}
where $\rho_{q}$, $\rho_{m}$  and  $H=\frac{\dot{a}}{a}$
are the density of new agegraphic dark energy, the pressureless matter density and the Hubble parameter,
respectively. We assume that there is no interaction between new agegraphic dark energy and
the pressureless matter, thus the continuity equation is given by
\begin{equation}
\dot{\rho_{q}}+3H\rho_{q}(1+\omega_{q})=0,\label{fgh}
\end{equation}
\begin{equation}
\dot{\rho_{m}}+3H\rho_{m}=0.\label{ggg}
\end{equation}
The fractional energy densities are also given by
\begin{equation}
\Omega_{q}=\frac{\rho_{q}}{3m_{p}^{2}H^{2}},\label{omeg}
\end{equation}
\begin{equation}
\Omega_{m}=\frac{\rho_{m}}{3m_{p}^{2}H^{2}}.\label{mmm}
\end{equation}
Using Eqs. (\ref{rhooo}) and (\ref{omeg}), we obtain
\begin{equation}
\Omega_{q}=\frac{n^{2}}{H^{2} \eta^{2}}.\label{omegg}
\end{equation}
Taking time derivative of Eq. (\ref{rhooo}) and using Eqs. (\ref{eta1}),
(\ref{fgh}), (\ref{omegg}) and $\dot{\eta}=\frac{1}{a}$, the new agegraphic dark energy Equation of State parameter (EoS) is obtained
\begin{equation}
\omega_{q}=-1+\frac{2\sqrt{\Omega_{q}}}{3na}.\label{eos}
\end{equation}
Using  $a=(1+z)^{-1}$, we can write Eq. (\ref{eos}) as follows
\begin{equation}
\omega_{q}=-1+\frac{2\sqrt{\Omega_{q}}(1+z)}{3n}.\label{eooos}
\end{equation}
Taking time derivative of Eq. (\ref{omegg}) and using $\dot{\eta}=\frac{1}{a}$
yields
\begin{equation}
\dot{\Omega_{q}}=-2H\Omega_{q}\Big(\frac{\dot{H}}{H^{2}}+\frac{\sqrt{\Omega_{q}}}{na}\Big).\label{dottt}
\end{equation}
Similarly, taking time derivative of Eq. (\ref{friedman}) and using Eqs. (\ref{fgh}),
(\ref{ggg}), (\ref{omeg}) and (\ref{mmm}) yields
\begin{equation}
\frac{\dot{H}}{H^{2}}=-\frac{3}{2}\Big(1+\Omega_{q}\omega_{q}\Big).\label{HH}
\end{equation}
Now, using Eqs. (\ref{eooos}),  (\ref{HH}) and inserting Eq. (\ref{dottt}), we obtain
\begin{equation}
\dot{\Omega_{q}}=3H\Omega_{q}(\Omega_{q}-1)(-1+\frac{2\sqrt{\Omega_{q}}(1+z)}{3n}).\label{dol}
\end{equation}
Using $\frac{d}{dt}=-(1+z)H\frac{d}{dz}$ and Eq.(\ref{dol}), one finds
\begin{equation}
\frac{d\Omega_{q}}{dz}=3\Omega_{q}(\Omega_{q}-1)(1+z)^{-1}(1-\frac{2\sqrt{\Omega_{q}}(1+z)}{3n}).\label{domegadz}
\end{equation}
The evolution of dimensionless Hubble parameter $E(z)=\frac{H}{H_{0}}$ in
new agegraphic  dark energy model is obtained by using Eqs. (\ref{eooos})
and (\ref{HH}) as follows
\begin{equation}
\frac{dE}{dz}=\Big(\frac{3E}{2(1+z)}\Big)\Big(1-\Omega_{q}+
\frac{2\Omega_{q}\sqrt{\Omega_{q}}(1+z)}{3n}\Big).\label{dedz}
\end{equation}
In figure (1), we have displayed the evolution of Equation of State parameter
$\omega_{q}$,
the evolution of density parameter $\Omega_{q}$ and the evolution of dimensionless Hubble parameter E(z) of new agegraphic dark energy model with respect to the redshift
parameter z. Also in figure (1), we have assumed the present values: $H_{0}=67.8~\frac{km}{s Mpc}$,
$\Omega_{m_{0}}=0.3$, $\Omega_{q_{0}}=0.7$ and $n=2.716$ \cite{weicai}.

\begin{figure}[ht]
\centering
The evolution of Equation of State parameter
of NADE model with respect to the redshift parameter $z$.
{\includegraphics[width=2in]{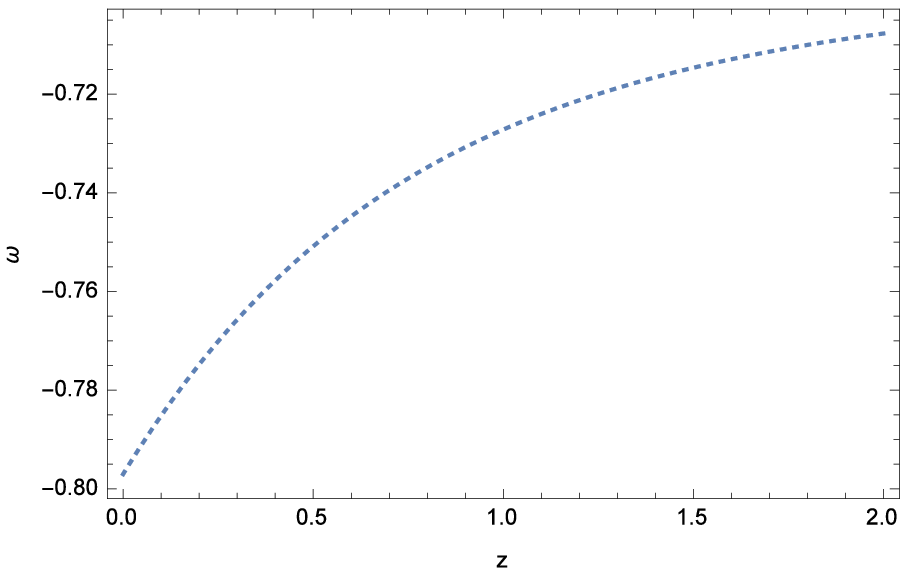}}

{The evolution of density parameter of NADE model with respect to the redshift parameter $z$.}
{\includegraphics[width=2in]{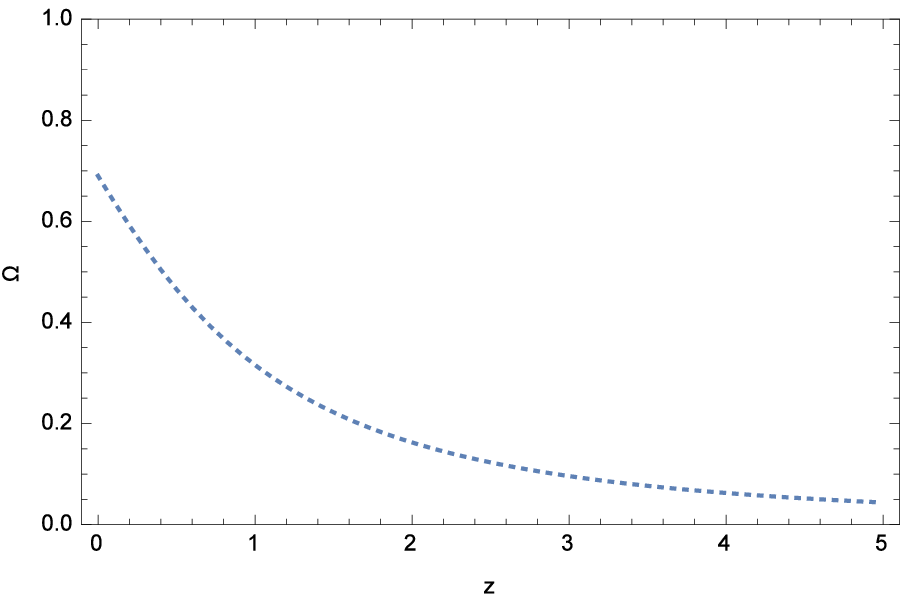}}

The evolution of dimensionless Hubble parameter in
 NADE model
and in the $\Lambda CDM$ model with respect to the redshift parameter $z$.
{\includegraphics[width=2in]{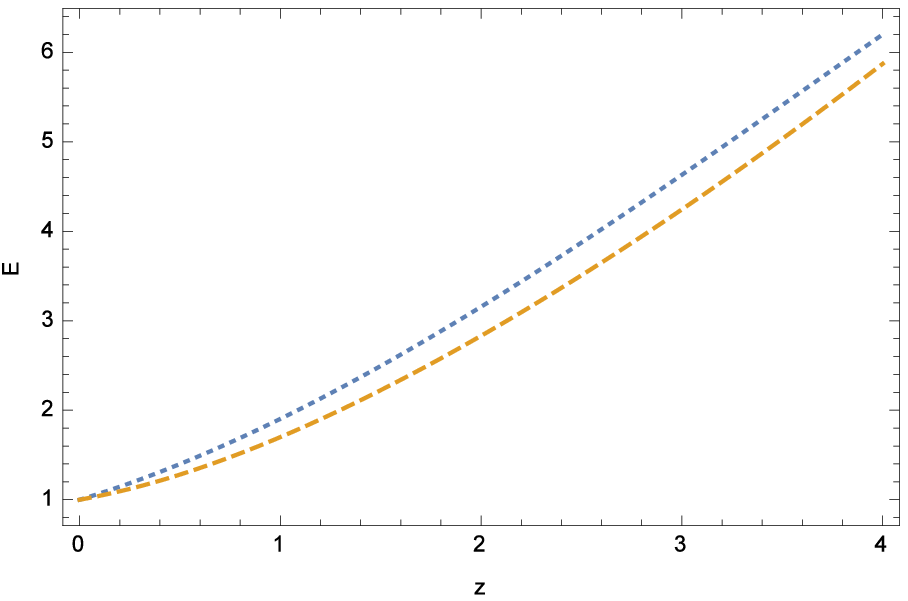}}
\caption{The dotted line represents the NADE model and the dashed line shows the  $\Lambda CDM$ model.}
\end{figure}
\section{Linear Perturbation Theory}
In this section, we discuss the linear perturbation theory of non-relativistic
dust matter, $g(a)$, for the new agegraphic dark energy model. Afterwards, we compare the new agegraphic  dark energy model with the EdS model and the $\Lambda CDM$ model. The differential equation for $g(a)$ is given by
\cite{perci,pace,fpace}
\begin{equation}
g''(a)+\Big(\frac{3}{a}+\frac{E'}{E}\Big)g'(a)-\frac{3}{2}\frac{\Omega_{m_{0}}}{a^{5}E^{2}}g(a)=0.\label{g}
\end{equation}
Using Eqs. (\ref{domegadz}) and (\ref{dedz}), we solve numerically Eq.
(\ref{g}) for studying the linear growth in new agegraphic dark energy model. Then, we compare
the linear growth in the new agegraphic dark energy model with the linear growths in the EdS model and the $\Lambda CDM$ model. Now, we plot the evolution
of $g(a)$ with respect to a function of the scale factor in figure (2). In the
new agegraphic dark energy model, the growth factor evolves more slowly compared to the $\Lambda
CDM$ model because the expansion of the universe slows down the structure formation. Also, in the $\Lambda CDM$ model, the growth factor evolves more slowly compared to the EdS model because the cosmological constant
dominates in the late time universe. These results are similar to the results obtained in the paper Malekjani \cite{holog} for holographic dark energy model.
\begin{figure}[ht]
  \centering
  \includegraphics[width=2in]{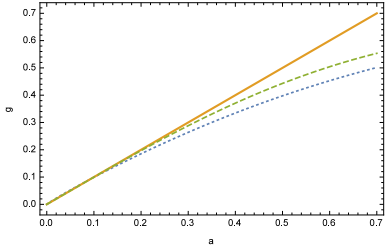}~
    \caption{Time evolution of the growth factor as a function of the scale
    factor. The dotted line indicates the  NADE model. The dashed
  line represent the  $\Lambda CDM$ model and the thick line shows the $EdS$ model.}
\end {figure}

\section{Spherical Collapse in the New Agegraphic of Dark Energy Model}
The discourse of structure formation  is obtained by the differential equation for the evolution of the matter perturbation $\delta$
in a matter dominated universe \cite{bernardf,Tpad}.
The differential equation for the evolution of $\delta$ in a universe including
a dark energy component was generalized in \cite{Abramo,lib}. Now, we consider the non- linear differential equation as given by \cite{pace}
\begin{equation}
\delta''+\Big(\frac{3}{a}+\frac{E'}{E}\Big)\delta'-\frac{4}{3}\frac{\delta'^{2}}{1+\delta}
-\frac{3}{2}\frac{\Omega_{m_{0}}}{a^{5}E^{2}}\delta=0, \label{nonlinear}
\end{equation}
where $'$ defines the derivative with respect to the scale factor $a$.
The linear differential equation for the evolution of $\delta$ is given by
\begin{equation}
\delta''+\Big(\frac{3}{a}+\frac{E'}{E}\Big)\delta'
-\frac{3}{2}\frac{\Omega_{m_{0}}}{a^{5}E^{2}}\delta=0. \label{linear}
\end{equation}
Now, in  Eqs. (\ref{nonlinear}) and (\ref{linear}) we consider the conditions
$\delta(10^{-4})=2.09\times10^{-4}$ and $\delta'(10^{-4})=0$ for the differential equation of perturbation in the EdS model \cite{pace}. In a similar way \cite{pace}, we obtain the conditions $\delta$ and $\delta'$  for the the new agegraphic dark energy and $\Lambda CDM$  models.

\begin{figure}[ht]
\centering
The linear growth of density perturbation in terms of
scale factor for different models
{\includegraphics[width=2in]{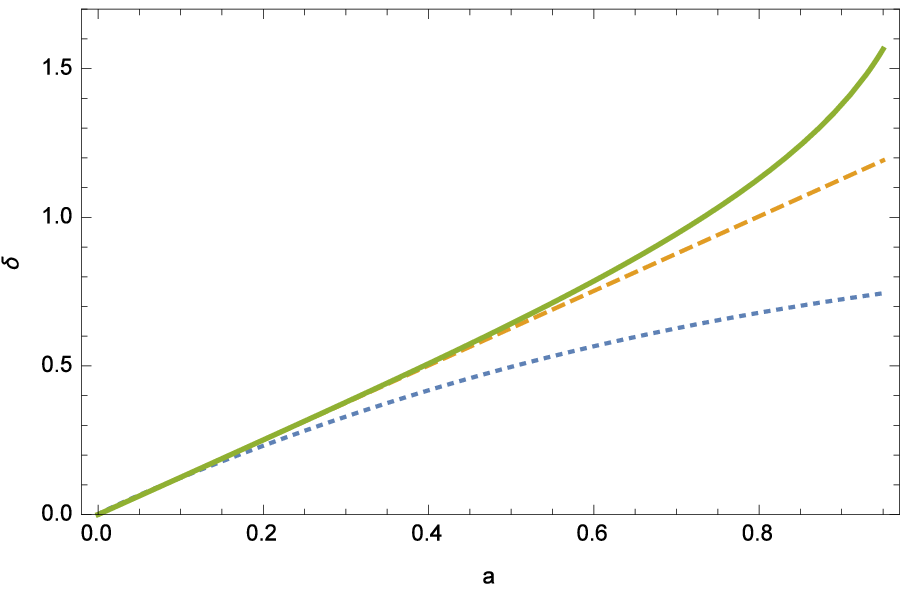}}

The non-linear growth of density perturbation in terms of scale factor for different models.
{\includegraphics[width=2in]{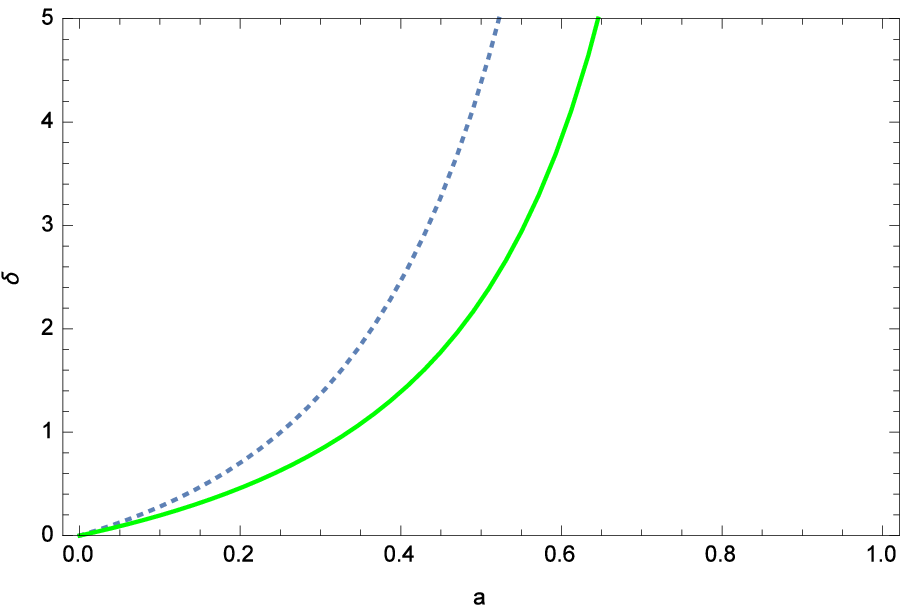}}
\caption{The dotted line represents the NADE model. The dashed line indicates the  $\Lambda CDM$ model and  the  thick line indicates the EdS model.}
\end{figure}

Figure (3-a) shows that in the new agegraphic dark energy model  the linear growth of density perturbation evolves more slowly compared to the $\Lambda CDM$ model and in the $\Lambda CDM$ model, the linear growth of density perturbation evolves more slowly compared to the EdS model. Also, figure (3-b) indicates that the non-linear growth of density perturbation in the new agegraphic dark energy model is faster than that of the EdS model.
\section{Determination of $\Delta_{vir}$ and $\delta_{c}$}
We consider the well known quantities of the spherical collapse model for the new agegraphic
 dark energy model: $\delta_{c}$ is the linear overdensity parameter,
the virial overdensity $\Delta_{vir}$ shows the halo size of structure, 
$\xi(z_{c})$ expresses the overdense spherical area of structure and  $\lambda(z_{c})$ represents the size structure. 
Now, we assume a spherical overdense region
with matter density $\rho$ in a surrounding universe defined by its background
dynamics with density $\rho_{b}$. The virial overdensity $\Delta_{vir}$ is
described by \cite{phys}
\begin{equation}
\Delta_{vir}=\frac{\rho}{\rho_{b}}\frac{R_{c}}{a_{c}},
\end{equation}
where $R_{c}$ is the virialization radius and $a_{c}$ is the scale factor corresponding
to virialization. Also, we can rewrite $\Delta_{vir}$  as follows \cite{phys}
\begin{equation}
\Delta_{vir}=1+\delta(a_{c})=\xi\Big(\frac{x_{c}}{\lambda}\Big)^{3},
\end{equation}
where
\begin{equation}
x_{c}=\frac{a_{c}}{a_{ta}},
\end{equation}
\begin{equation}
\xi=\frac{\rho(R_{ta})}{\rho_{b}(a_{ta})}=1+\delta(a_{ta}).
\end{equation}
Here, $R_{ta}$ is the turn-around radius and $a_{ta}$ is the scale factor corresponding to the turn-around epoch. Also, we use the virial radius $\lambda$
as follows \cite{wW}
\begin{equation}
\lambda=\frac{1-\frac{\eta_{\nu}}{2}}{2+\eta_{t}-\frac{3}{2}\eta_{\nu}},
\end{equation}
where
\begin{equation}
\eta_{\nu}=\frac{2}{\xi}\frac{\Omega_{q}(a_{c})}{\Omega_{m}(a_{c})}\Big(\frac{a_{ta}}{a_{c}}\Big),
\end{equation}
\begin{equation}
\eta_{t}=\frac{2}{\xi}\frac{\Omega_{q}(a_{ta})}{\Omega_{m}(a_{ta})}.
\end{equation}
Here, $\eta_{\nu}$ and $\eta_{t}$ are the (Wang-Steinhardt) WS parameters.
Now, we discuss the results obtained for $\delta_{c}(z_{c})$, $\lambda(z_{c})$, $\xi(z_{c})$ and $\Delta_{vir}(z_{c})$ in the models introduced in this
paper.

In the figure (4), we see that in the EdS model, $\delta_{c}=1.686$  and it is independent of the redshift $z_{c}$.
In the $\Lambda CDM$ model, $\delta_{c}$  is smaller than 1.686 and its value is approximately the same as that  of  EdS model at high redshifts.
Therefore the universe is matter dominated at high redshift and the cosmological constant dominates at low redshift. We can state that the primary structures form with a lower critical density. 
 Also, in the new agegraphic dark energy model, $\delta_{c}$ is smaller than that of  $\Lambda CDM$ model. This is due to the fact that in figure (1c) the Hubble parameter in the new agegraphic dark energy model is larger than that of the $\Lambda CDM$ model. Hence, there is a dark energy dominated universe at low redshifts and there is a matter
dominated universe at high redshifts.

In the figure (5), we see that in the EdS model, $\lambda(z_{c})=0.5$  and it is independent of the redshift $z_{c}$.
In the $\Lambda CDM$ model, $\lambda(z_{c})$  is smaller than 0.5 and its value is approximately the same as that of  EdS model at high redshifts . Also, in the new agegraphic dark energy model, $\lambda(z_{c})$ is smaller than that of  $\Lambda CDM$ model. Thus, we find that the size of structures in the new agegraphic dark energy model is smaller than that of the $\Lambda CDM$ model.

\begin{figure}[ht]
  \centering
  \includegraphics[width=2in]{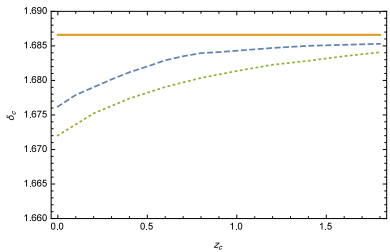}~
    \caption{The time evolution of  linear overdensity, $\delta_{c}(z_{c})$,
in terms of a function of the collapse redshift for the 
NADE model,
the $\Lambda CDM$ model, and the EdS model.
The dotted line represents the NADE model. The dashed line indicates the  $\Lambda CDM$ model and  the thick line indicates the EdS model.}
\end {figure}

\begin{figure}[ht]
\centering
\includegraphics[width=2in]{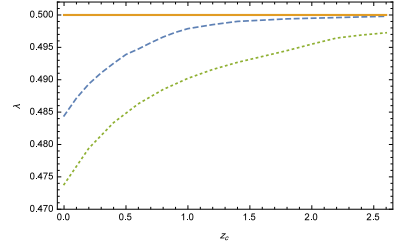}~
\caption{The virial radius $\lambda(z_{c})$ in terms of the collapse redshift
$z_{c}$ for the NADE model, the $\Lambda CDM$ model and the EdS model.
The dotted line represents the NADE model. The dashed line indicates the $\Lambda CDM$ model and  the  thick line indicates the EdS model.}
\end {figure}

\begin{figure}[ht]
\centering
The variation of $\xi(z_{c})-z_{c}$ for the NADE model, the $\Lambda CDM$ model and the EdS model.
{\includegraphics[width=2in]{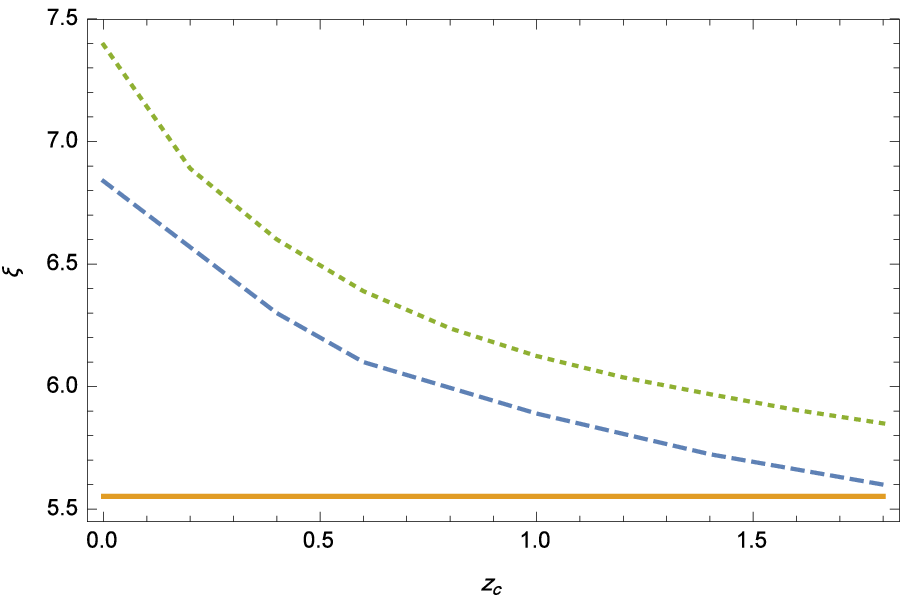}}

The variation of $\Delta_{vir}(z_{c})-z_{c}$ for the 
NADE model, the $\Lambda CDM$ model and the EdS model.
{\includegraphics[width=2in]{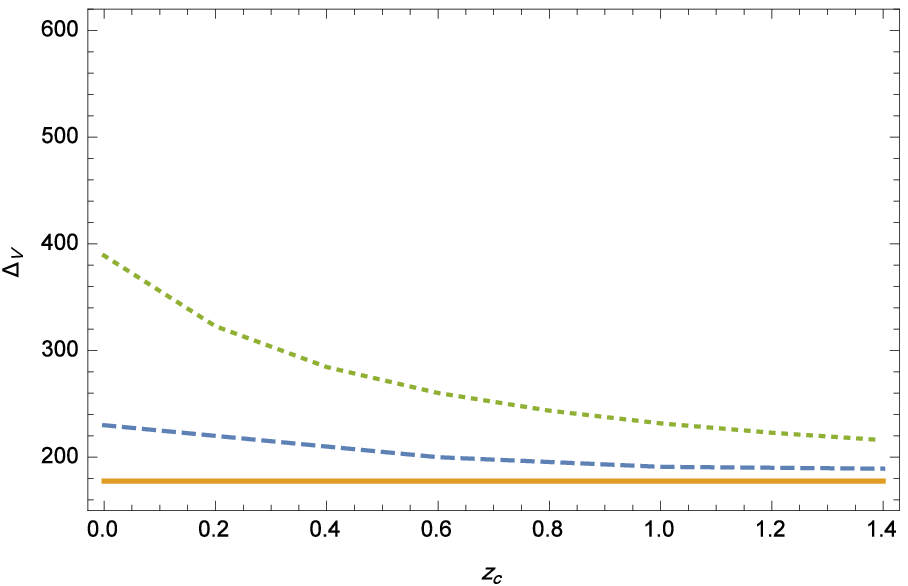}}
\caption{The dotted line represents the  NADE model. The dashed line indicates the  $\Lambda CDM$ model and  the  thick line indicates the EdS model.}
\end{figure}

In the figure (6a), we see that in the EdS model, $\xi(z_{c})=5.6$  and it is independent of the redshift $z_{c}$.
In the $\Lambda CDM$ model, $\xi(z_{c})$  is larger than 5.6 but its value is approximately the same as that of  EdS model at high redshifts. Also, in the new agegraphic dark energy model, $\xi(z_{c})$ is larger than the $\Lambda CDM$ model. Thus, we find that in the new agegraphic dark energy model, the overdense spherical area is denser than the EdS model and the $\Lambda CDM$ model.

In the figure (6b), we see that in the EdS model, $\Delta_{vir}(z_{c})=178$  and it is independent of the redshift $z_{c}$.
In the $\Lambda CDM$ model, $\Delta_{vir}(z_{c})$  is larger than 178 but its value is approximately the same as that of  EdS model at high redshifts. Also, in the new agegraphic dark energy model, $\Delta_{vir}(z_{c})$ value is larger than the $\Lambda CDM$ model. Thus, we find that in the new agegraphic dark energy model, the halo size is larger than those of  EdS and  $\Lambda CDM$ models.
\section{ mass function and number density}
In this section,  we calculate the number density and the mass function in
a given mass range. The average comoving number density of halos of mass
$M$ is described by \cite{phw,cole}
\begin{equation}
n(M,z)=\Big(\frac{\rho}{M^{2}}\Big)
\Big(\frac{d \log \nu}{d \log M}\Big)\nu f(\nu).\label{nmz}
\end{equation}
Here, $\rho$ is the background density and $f(\nu)$ is the multiplicity function.
Also, $\nu$ is described by
\begin{equation}
\nu=\frac{\delta_{c}^{2}}{\sigma^{2}(M)},
\end{equation}
where $\sigma(M)$ is the r.m.s of the mass fluctuation in the sphere of mass
M. The formula $\sigma(M,z)$ is given by \cite{ptpv}
\begin{equation}
\sigma(M,z)=\sigma_{8}(z)\Big(\frac{M}{M_{8}}\Big)^{-\frac{\gamma(M)}{3}}.\label{sigmam}
\end{equation}
Here, $\sigma_{8}$ and $M_{8}$ are the mass variance of the overdensity
on the scale of $R_{8}$ and mass inside a sphere, respectively. $R_{8}$ is the radius inside a sphere. The numerical values $R_{8}$ and $M_{8}$ are $8h^{-1}Mpc$
and $6\times10^{14}\Omega_{m_{0}}h^{-1}M_{\odot}$, respectively.
 The formula $\sigma_{8}(z)$ is given by \cite{ptpv}
\begin{equation}
\sigma_{8}(z)=g(z)\sigma_{8}(M,z=0),\label{sigma8}
\end{equation}
where $g(z)$ is the linear growth factor. The formula $\sigma_{8,DE}(M,z=0)$ is given by
\begin{equation}
\sigma_{8,DE}(M,z=0)=0.8\Big[\frac{\delta_{c,DE}(z=0)}
{\delta_{c,\Lambda CDM}(z=0)}\Big].\label{sig}
\end{equation}
The formula $\gamma(M)$ is described by
\begin{equation}
\gamma(M)=\Big(0.3\Gamma+0.2\Big)\Big[2.92+\frac{1}{3}\log(\frac{M}{M_{8}})\Big],
\end{equation}
where
\begin{equation}
\Gamma=\Omega_{m_{0}}h \exp(-\Omega_{b}-\frac{\Omega_{b}}{\Omega_{m_{0}}}).
\end{equation}
Eqs. (\ref{sigmam}), (\ref{sigma8}) and (\ref{sig}) have validity limits
\cite{ptpv}. They represent that the fitting formula predicts lower values
of the values of the variance for $M>M_{8}$ and the fitting formula predicts higher values of the values of the variance for $M<M_{8}$. Now, we can use the ST mass function formula given by \cite{rksheth,tormen}
\begin{equation}
\nu f_{ST}(\nu)=A_{1}\sqrt{\frac{b\nu}{2\pi}}\Big[1+\frac{1}{(b\nu)^{p}}\Big]\exp(-\frac{b\nu}{2}).\label{fst}
\end{equation}
Here the numerical parameters are: $A_{1}=0.3222$, $p=0.3$ and $b=0.707$.
Also, we use the PO mass function formula given by \cite{adel,apop}
\begin{eqnarray}
\nu f_{PO}(\nu)=A_{2}\Big[1+\frac{0.1218}{(b\nu)^{0.585}}+\frac{0.0079}{(b\nu)^{0.4}}\Big]
\sqrt{\frac{b\nu}{2\pi}}\nonumber\\
\exp\Big(-0.4019b\nu
\Big[1+\frac{0.5526}{(b\nu)^{0.585}}+\frac{0.02}{(b\nu)^{0.4}}\Big]^{2}\Big),\label{fpo}
\end{eqnarray}
where the numerical parameter is: $A_{2}=1.75$. The YNY mass function formula is presented by \cite{yahagi}
\begin{equation}
\nu f_{YNY}(\nu)=A_{3} \nu^{\frac{D}{2}} \Big[1+\Big(B\sqrt{\frac{\nu}{2}}\Big)^{C}\Big]
\exp\Big[(-B\sqrt{\frac{\nu}{2}})^{2}\Big],\label{fyny}
\end{equation}
where the numerical parameters are: $A_{3}=0.298$, $B=0.893$, $C=1.39$ and
$D=0.408$.

\begin{figure}[ht]
\centering
The evolution of the mass function for the new agegraphic dark energy and the $\Lambda CDM$ models in the case $z=0$.
{\includegraphics[width=2in]{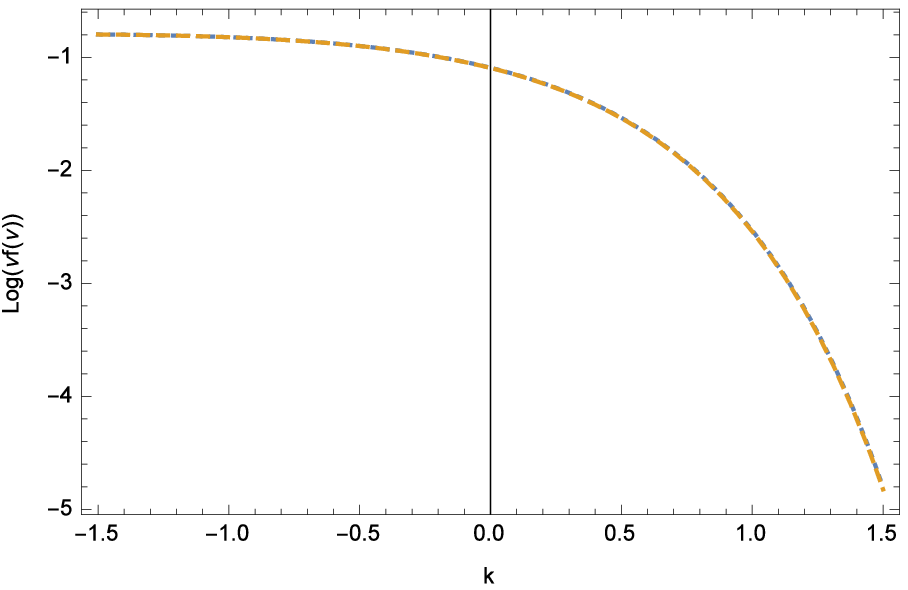}}

The evolution of the mass function
 for the NADE model and the $\Lambda CDM$ model in the case $z=1$.
\label{ihbii}{\includegraphics[width=2in]{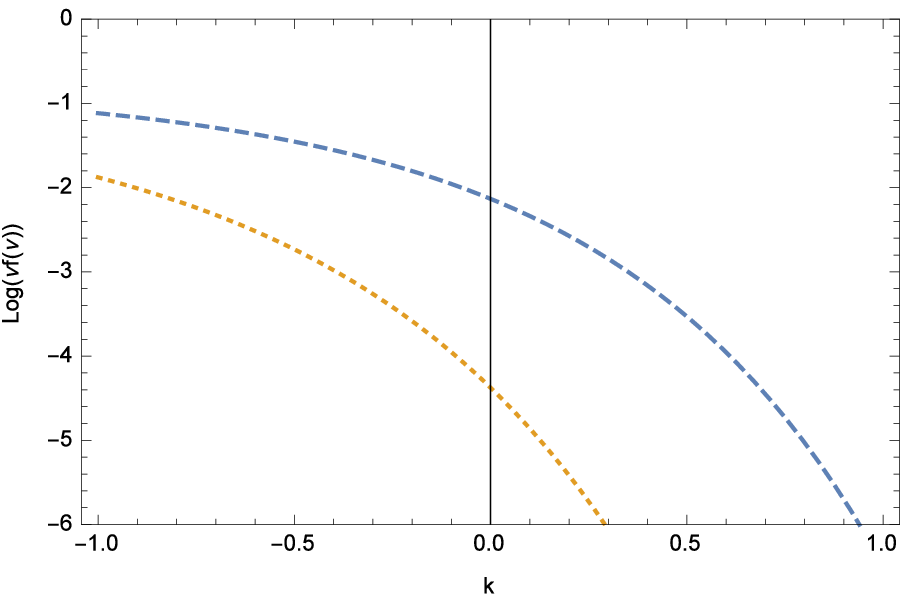}}
\caption{The dotted line represents the NADE model and the dashed  line indicates the  $\Lambda CDM$ model.}
\end{figure}

\begin{figure}[ht]
\centering
The evolution of the number density
 for NADE model and  the $\Lambda CDM$ model in the case $z=0$.
 {\includegraphics[width=2in]{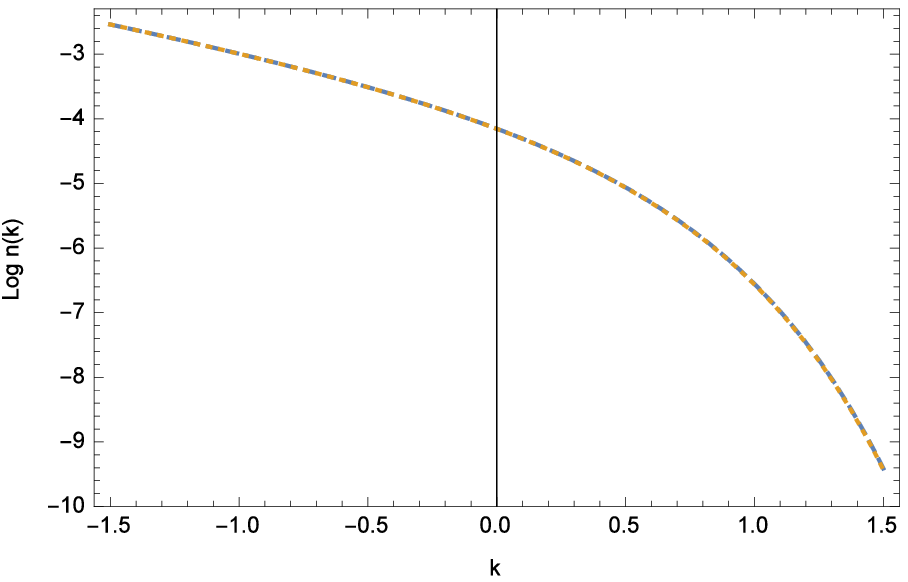}}

The evolution of the number density for the new agegraphic dark energy model and the $\Lambda CDM$ model in the case $z=1$.
{\includegraphics[width=2in]{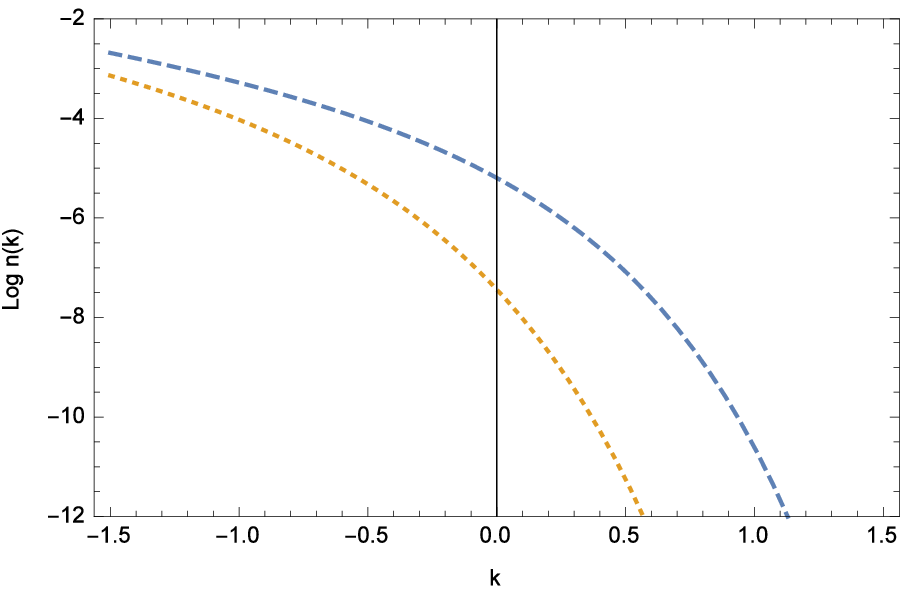}}
\caption{The dotted line represents the NADE model and the dashed line indicates the  $\Lambda CDM$ model.}
\end{figure}
\begin{figure}[ht]
\centering{\includegraphics[width=2in]{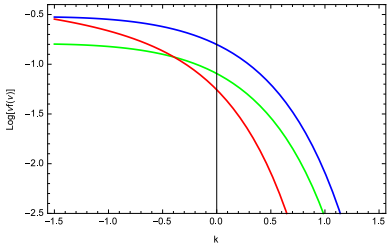}}
\caption{ The evolution of the various mass functions with respect to $k$ for the NADE model in the case
$z=0$. The green line represents ST mass function, the blue line represents PO mass function and the red line represents YNY mass function }
\end{figure}

Now, we discuss the evolution of the ST mass function with represent to $k$
for the new agegraphic dark energy model and the $\Lambda CDM$ model in the figure (7). The $k$ formula is defined as $k=\log \frac{M}{M_{8}}$. In the figure (7), the evolution of the ST mass function with represent to $k$ is identical for the new agegraphic dark energy model and the $\Lambda CDM$ model in the $z=0$ case, but it is different for the new agegraphic dark energy model and the $\Lambda CDM$ model in the $z=1$ case. This difference is due to the difference between   $g(z)$ and $\delta_{c}$ in
two models. Also, $g(z)$ and $\delta_{c}$  are dependent on the redshift.
Using Eqs. (\ref{nmz}) and (\ref{fst}), we obtain the average comoving number
density of halos of mass $M$ for the new agegraphic dark energy model and the $\Lambda CDM$ model
in the  $z=0$ and $z=1$ cases. In the figure (8), the evolution of the number
density with represent to $k$ is identical to those of the new agegraphic dark energy model and the $\Lambda CDM$ model in the $z=0$ case, but it is different for the new agegraphic dark energy model and the $\Lambda CDM$ model in the $z=1$ case. In the figure (8b), for small objects the difference in the number densities of halo objects is low but the difference in the number densities of halo objects is increasing for high mass in the new agegraphic dark energy model and the $\Lambda CDM$ model. Therefore, we find
that the number of objects per unit mass is increasing for high mass in
the new agegraphic dark energy model and the $\Lambda CDM$ model.
Now, using Eqs. (\ref{fst}), (\ref{fpo}) and (\ref{fyny}), we compare the
various mass functions at $k$ in figure (9). We can see that the PO mass function
is larger than YNY mass function and ST mass function, for all mass scales.
{\section{Comparison between the new agegraphic dark energy model
with the tachyon scalar field and the holographic dark energy models}}
In this section, we express the results of the evolution of spherical overdensities in new agegraphic dark energy model and compare
our results with the results of the tachyon scalar field
model (for all $n$) \cite{tachyon} and the holographic dark energy model
(only for $c=0.815$)\cite{holog}.

In the new agegraphic dark energy model, the growth factor evolves more slowly compared to the $\Lambda CDM$ model because the expansion of the universe slows down the structure formation. Also, in the $\Lambda CDM$ model, the growth factor  evolves more slowly compared to the EdS model because the cosmological constant dominates in the late time universe.
In the tachyon scalar field model, at the beginning, the growth factor is larger than the EdS and the $\Lambda CDM$ models for small scale factors,
but for larger scale factors, its growth factor is smaller than the EdS model while it is still larger than the $\Lambda CDM$ model. Therefore,
at first, the tachyon scalar field model predicts the structure formation
more impressive than the EdS and the $\Lambda CDM$ models and over time,
the structure formation in the tachyon scalar field model coincides with the EdS and the $\Lambda CDM$ models \cite{tachyon}. The structure formation
in the holographic
dark energy model  is similar to  the new agegraphic dark energy model \cite{holog}.

In the new agegraphic dark energy model, the linear overdensity parameter $\delta_{c}$  is larger than the linear overdensity parameters in the tachyon scalar field model and the holographic dark energy model, respectively. This means that the Hubble parameter in the new agegraphic dark energy model is smaller than the hubble parameter in the tachyon scalar field model and the holographic dark energy model, respectively.

We may compare $\lambda(z_{c})$ for the new agegraphic dark energy
model, the tachyon scalar field model and the holographic dark energy model. We find that the size of structures in the holographic dark energy model is larger than those of the new agegraphic dark energy and the tachyon scalar field models.

Also, we can conclude that in the tachyon scalar field model, $\xi(z_{c})$ is denser than the new agegraphic dark energy model and the holographic dark energy model, respectively. We can also claim that in the tachyon scalar field model, the halo size is larger than those of the new agegraphic dark energy model and the holographic dark energy model.

Finally, we discuss the evolution of the ST mass function with represent to $k$ for the new agegraphic dark energy model, the tachyon scalar field model and the holographic dark energy model in the $z=0$ and $z=1$ cases. The evolution of the ST mass function with represent to $k$ is the same for the three models described above in the $z=0$ case but it is different from them in the $z = 1$ case. Therefore, the evolution of the ST mass function in the new agegraphic dark energy model is smaller than those of the holographic and the tachyon dark energy models in the z=1 case, respectively.

Also, we compare the average comoving number density of halos of mass $M$ for the new agegraphic, the tachyon and the holographic dark energy models in the $z=0$ and $z=1$ cases.
We can claim that the evolution of the number density with represent to $k$ is identical for the new agegraphic, the tachyon and the holographic dark energy models in the $z=0$ case, but it is different from them in the $z=1$ case. The evolution of the number density in the new agegraphic dark energy model is smaller than those of the holographic and the tachyon dark energy models in the $z = 1$ case.
Thus, we can claim that the number of objects per unit mass increases for high mass in the new agegraphic, the holographic and the tachyon dark energy models, respectively.
We compare the various mass functions at $k$ for the new agegraphic, the tachyon and the holographic dark energy models in the $z=0$. We can see
that the PO mass function is larger than YNY mass function and ST mass function for the three models described above.

\section{Concluding remarks}\label{Con}

In this work, we discussed the evolution of spherical overdensities
in the new agegraphic dark energy model. We obtained the evolution of the dimensionless Hubble parameter $E(z)$, the evolution of density parameter $\Omega_{q}$ and the evolution of the equation of state parameter $\omega_{q}$
for the new agegraphic dark energy model with respect to the cosmic redshift function. We compared the linear growth in the new agegraphic  dark energy model with the linear growth in the EdS model and the $\Lambda CDM$ model: In the new agegraphic dark energy model, the growth factor evolves more slowly compared to the $\Lambda CDM$ model because the expansion of the universe slows down the
structure formation. Also, in the $\Lambda CDM$ model, the growth factor  evolves more slowly compared to the EdS model because the cosmological constant dominates in the late time universe.

We showed that in the EdS model, $\delta_{c}$ is independent of the redshift $z_{c}$ and in the new agegraphic dark energy model, $\delta_{c}$ is smaller than that of the $\Lambda CDM$ model  because in figure (1c) the Hubble parameter in the new agegraphic dark energy model is larger than that of the $\Lambda CDM$ model. Hence, there is a dark energy dominated universe at low redshifts and there is a matter
dominated universe at high redshifts.

We saw that in the EdS model, $\lambda(z_{c})$ is independent of the redshift $z_{c}$ and the size of structures in the new agegraphic dark energy model is smaller than that
of the $\Lambda CDM$ model.
Also, we concluded that in the EdS model, $\xi(z_{c})$ is independent of the redshift $z_{c}$ and the overdense spherical area  in the new agegraphic dark energy model is denser than those of the EdS model and the $\Lambda CDM$ model.
We found that in the EdS model, $\Delta_{vir}(z_{c})$ is independent of the redshift $z_{c}$ and in the new agegraphic dark energy model the halo size is larger than those
of the EdS model and the $\Lambda CDM$ model.

Finally, we discussed  the evolution of the ST mass function with represent to $k$ for the new agegraphic dark energy model and the $\Lambda CDM$ model. We saw that the evolution of the ST mass function with represent to $k$ is identical to the new agegraphic dark energy model and the $\Lambda CDM$ model in the $z=0$ case, but it is different from the new agegraphic dark energy model and the $\Lambda CDM$ model in the $z=1$ case.
We studied the average comoving number density of halos of mass $M$ for the new agegraphic dark energy model and the $\Lambda CDM$ model in the $z=0$ and $z=1$ cases . We saw that  the evolution of the number density with represent to $k$ is identical for the new agegraphic dark energy model and the $\Lambda CDM$ model in the $z=0$ case, but it is different from the new agegraphic dark energy model and the $\Lambda CDM$ model in the $z=1$ case. In the figure (8b), for small objects the difference in the number densities of halo objects is low but the difference in the number densities of halo objects is increasing for high mass in the new agegraphic dark energy model and the $\Lambda CDM$ model. Therefore, we found that the number of objects per unit mass
is increasing for high mass in the new agegraphic dark energy model and the $\Lambda CDM$ model.
Moreover, we compared the results of the evolution of spherical overdensities in the new agegraphic  dark energy model with the results of the tachyon scalar field model (for all n) \cite{tachyon} and the holographic dark energy model (only for $c=0.815$) \cite{holog}.


\end{document}